\documentstyle[11pt,aaspp4]{article}

\lefthead{Tsuneta \& Naito}

\righthead{Fermi acceleration at fast shock in a solar flare
and impulsive loop-top hard X-ray source}

\begin{document}

\title{FERMI ACCELERATION AT FAST SHOCK IN A SOLAR FLARE
AND IMPULSIVE LOOP-TOP HARD X-RAY SOURCE}

\author{Saku Tsuneta}
\affil{National Astronomical Observatory, Mitaka, Tokyo 181}

\author{Tsuguya Naito}
\affil{Department of Earth and Planetary Physics, University of Tokyo,
Hongo, Bunkyo-ku, Tokyo 113}

\begin{abstract}

Fermi acceleration has not been considered to be viable to explain
non-thermal electrons (20$\sim$100 KeV) produced in solar flares,
because of its high injection energy.  Here, we propose that
non-thermal electrons are efficiently accelerated by first-order Fermi
process at the fast shock, as a natural consequence of the new
magnetohydrodynamic picture of the flaring region revealed with {\it
Yohkoh}. An oblique fast shock is naturally formed below the
reconnection site, and boosts the acceleration to significantly
decrease the injection energy. The slow shocks attached to the
reconnection X-point heat the plasma up to 10$\sim$20 MK, exceeding
the injection energy.  The combination of the oblique shock
configuration and the pre-heating by the slow shock allows bulk
electron acceleration from the thermal pool.  The accelerated
electrons are trapped between the two slow shocks due to the magnetic
mirror downstream of the fast shock, thus explaining the impulsive
loop-top hard X-ray source discovered with {\it Yohkoh}.  Acceleration
time scale is $\sim$ 0.3--0.6 s, which is consistent with the time
scale of impulsive bursts.  When these electrons stream away from the
region enclosed by the fast shock and the slow shocks, they are
released toward the footpoints and may form the simultaneous
double-source hard X-ray structure at the footpoints of the
reconnected field lines.

\end{abstract}


\section{INTRODUCTION}

Hard X-ray spectral observations of solar flares made in past 20 years
have established that efficient electron acceleration
($\sim10^{34-35}$ electrons s$^{-1}$) occurs during impulsive solar
flares (Lin \& Hudson 1971). The {\it Yohkoh} Hard X-ray Telescope
clearly showed that almost all flares have double footpoint sources,
the signature of bombardment by the accelerated electrons into the
dense chromosphere (Hoyng et al. 1981, Sakao et al. 1997a, 1997b).
These spectral and imaging observations showed that rich amount of
non-thermal electrons were accelerated somewhere in the system, and
were injected in the soft X-ray loop, which turned out to be the
reconnected field lines filled with evaporated plasma (Tsuneta 1996).
Furthermore, a surprising discovery from the HXT is the detection of
an impulsive hard X-ray source located above the soft X-ray loop
(Masuda et al. 1994). The overall time profile of the loop-top source
is similar to those from the footpoints, implying common origin of
acceleration.  See Miller et al. (1997) for a review of these recent
observations and the theoretical implications.

{\it Yohkoh} soft X-ray observations, on the other hand, strongly
suggest that magnetic reconnection serves as an efficient engine to
convert magnetic energy to plasma kinetic and thermal energies
(Tsuneta 1996). In this process, MHD slow-mode shocks attached to the
reconnection X-point appear to convert magnetic energy, brought with
the inflow to the X-point from the vast active region corona, to
plasma heating and an outflow jet with the Alfv\'en speed on the
upstream side. The outflow jet can form a fast-mode shock
(quasi-perpendicular shock) to further heat the plasma (Masuda et al.
1994, Shibata et al.  1995, Tsuneta et al. 1997).

The observed geometry indicated that the loop-top hard X-ray source
was located at the downstream side of reconnection site (below the
X-point).  This implied that the formation of the loop-top source was
related to the fast shock, regardless of whether it was of thermal or
non-thermal origin (Tsuneta et al. 1997).  Furthermore, the electron
spectra of the loop-top hard X-ray source turned out to be more
consistent with the power-law with differential power index of 2--4
(Alexander \& Metcalf 1997). Aschwanden et al. (1996) concluded from
their innovative time-of-flight method that the acceleration site of
this flare must have been located around the loop-top hard X-ray
source or above, if the acceleration site was spatially concentrated.
All these three independent approaches point to the loop-top hard
X-ray source as the acceleration site.

In this Letter, we suggest that the electron acceleration and the
resultant formation of the loop-top hard X-ray source are natural
consequences of the new MHD framework being established by the {\it
Yohkoh} soft X-ray observations: The highly oblique fast shock
sandwiched by the slow shocks provides an ideal site for electron
acceleration with the first-order shock Fermi acceleration (eg.
Blandford \& Eichler 1987; Jones \& Ellison, 1991).

\section{SHOCK FERMI ACCELERATION}

\subsection{Requirements from Observations}

There are four stringent requirements from the observations to
quantitatively examine whether first-order Fermi acceleration is
viable in the 10--100 keV electron acceleration.

(1) Since the background plasma density is high in the solar corona,
the acceleration rate has to exceed the collisional loss at the
thermal energy of the background plasma. The net acceleration rate is
given by
\begin{equation}
\frac{dE}{dt} = \frac{dE}{dt}_{\rm a} - \left|\frac{dE}{dt}_{\rm c}\right|,
\end{equation}
where the first term on the right hand side is the acceleration rate
and the second term the collisional energy loss. The net acceleration
rate has to be positive at the thermal energy of the background
plasma.  (2) The energy gain has to be high enough to explain the
loop-top hard X-ray source, that is observed at $\sim$ 50 keV. (If the
loop-top source provides electrons responsible for the footpoint hard
X-rays, the energy has to reach $>$100 keV.) Electrons, therefore,
have to be accelerated within a relatively short time, over which an
oblique (quasi-perpendicular) field line crosses the fast shock
structure.  (3) Acceleration takes place on a time scale of impulsive
bursts, which is about 1s or faster. (4) Number of accelerated
electrons is consistent with number of accelerated electrons deduced
from observations (10$^{34-35}$ electrons s$^{-1}$.)

\subsection{Injection Energy and Energy Gain}

In this paper, we call the energy, at which the net energy gain rate
[Eq. (1)] equals zero, the injection energy.  Although the injection
energy is defined with regard both to energy loss and particle
scattering, we assume that there exists sufficient magnetic field
turbulence such as whistler waves to be able to scatter the electrons
with energies exceeding 1 keV (eg. Melrose 1986, Melrose 1994). [These
waves can energize electrons (eg. Miller et al. 1997). We, however, do
not include these alternative acceleration processes such as resonance
acceleration and DC electric acceleration in this analysis.]
  
The collisional energy loss in Eq.(1) is estimated to be
\begin{equation}
\left|\frac{dE}{dt}_{\rm c}\right| \sim 47 \frac{n_{\rm 10}}{\surd E},
\end{equation}
where $n_{\rm 10}$ is the background electron density in unit of
10$^{10}$ cm$^{-3}$, and $E$ the electron energy in keV (eg. Jackson
1975).  In the scheme of shock acceleration, the acceleration rate is
derived from energy gain $\Delta E$ and required time $\Delta t$ per
each shock crossing, and represented by
\begin{equation}
\frac{dE}{dt}_{\rm a} = \frac{\Delta E}{\Delta t}
= \frac{2}{3} E \frac{u}{l \cos\theta},
\end{equation}
where $l$ is the (energy-independent) diffusion length measured along
the field line, $u$ the upstream speed with respect to the downstream
speed, and $\theta$ the angle between the fast shock normal and the
field line crossing the shock(Fig. 1).  Here, we apply $\Delta E =
8uE/3v\cos\theta$, where $v$ is an electron velocity ($E=mv^2/2$), and
$1/\cos\theta$ is the enhancement factor originating from the
obliqueness of the shock.  Assuming an isotropic electron
distribution, we apply $\Delta t = l/(v/4)$.  In the standard shock
acceleration theory, $l = \kappa_{\rm 1}/u_{\rm 1}+\kappa_{\rm
2}/u_{\rm 2}$, where $\kappa_{\rm 1 (2)}$ and $u_{\rm 1 (2)}$ are the
diffusion coefficient and flow speed in the upstream (downstream),
respectively.  The time during which a particular field line crosses
the fast shock is $L/2u\tan\theta$. The crossing time decreases with
increasing obliqueness. From Eq. (3), the energy gain is given by
\begin{equation}
\frac{E}{E_{\rm 0}} = \exp ( \frac{L}{3 l \sin\theta}),
\end{equation}
for particles with energy above the injection energy.  

We simply represent the diffusion length as a parameter $l$.  We adopt
$\theta$ as another parameter expressing obliqueness of the shock.  It
is proposed that the efficiency of shock acceleration is boosted due
to the obliqueness (Jokipii 1987, Naito \& Takahara 1995a, 1995b and
references therein).  Fig. 2 shows that the injection energy rapidly
decreases with increasing obliqueness.  Here, we adopt the following
values from the observations; the density of the background plasma $n
= 10^{10}$ cm$^{-3}$, the temperature $T = 2 \times 10^{7}$ K (Tsuneta
et al. 1997), the length of the fast shock $L = 7000$ km, which is the
width of the loop-top hard X-ray source (Masuda et al. 1984). We
assume the diffusion length $l = 500 {\rm km}$, which will be
estimated later to satisfy the observational and theoretical
requirements. The speed of the outflow from the reconnection region is
equal to the Alfv\'en velocity of the upstream side (Petschek 1964),
and is taken to be $u = 1000$ km s$^{-1}$ (Tsuneta 1996).

\subsection{Diffusion Length and Shock Angle}

The number of accelerated electrons critically depends on the
injection energy with respect to the thermal energy of the background
plasma. Here, we conservatively assume that the injection energy is
equal to the thermal energy of the background plasma heated by the
slow shocks ($\sim$ 2 keV), implying bulk electron acceleration,
though higher injection energy resulting in acceleration of tail
electrons in the thermal distribution function may be acceptable.  The
initial energy of electrons $E_{\rm 0}$ is, therefore, 2 keV, which is
accepted as the injection energy for $\theta = 85$ deg as shown in
Fig. 2.  Combining with the electron energy at the loop-top hard X-ray
source of about 100 keV, we assume that the energy gain of 50 or more
is necessary.

Since the diffusion (scattering) length $l$ and the shock angle
$\theta$ are parameters that cannot be directly determined from the
observations, we obtain the two parameters from the following two
conditions to satisfy the first two requirements in section 2.1;
\begin{equation}
\frac{dE}{dt} > 0 \;\;\;{\rm at}\;\;\; E_{\rm 0} = 2 {\rm keV},
\end{equation}
\begin{equation}
\frac{E}{E_{\rm 0}} > 50.
\end{equation}
Fig. 3 shows $dE/dt=0$ at $E_{\rm 0}=2$ and $1$KeV, and $E/E_{\rm
0}=25,50$ and $100$ in the $\theta - l$ space.  The following two
additional conditions have to be satisfied to choose the proper
diffusion length $l$: (1) The length must be comparable to or be
larger than the diffusion length derived from the assumption of the
Bohm diffusion (eg. Kirk 1994).  It is customarily expressed by using
Bohm diffusion coefficient $ \kappa_{\rm B} $ as $l \sim \eta
\kappa_{\rm B}/u \sim 2 \eta E/(3 m \omega_{\rm ce} u) \sim 0.06 \eta $ km
for $E=100$keV and $B=10$G where the factor $\eta $ is anticipated to
be in the range of $\eta \sim 1-100 $ from recent observations in
interplanetary shocks (Kang \& Jones 1997; Baring et al. 1997).  Since
scattering by whistler waves in transient phenomena such as solar
flares is poorly understood, we conservatively assume $\eta \sim
10^{4} $, and the diffusion length $l = 600 {\rm km}$.  (2) The length
must be smaller than the distance between the slow shocks ($\sim L$),
which contain the electrons, thus $l \leq L/2\sin\theta \sim
3500/\sin\theta$ km.

These conditions given above are met when the shock is highly oblique
($\theta \sim 80-85$ degree). The diffusion length in this case is
about $l \sim 600$ km. The diffusion length $l$ needs to be shorter
for higher energy gain. For instance, $l\sim 500$ km and $\theta\sim
85$ degree are needed to accelerated electrons to 1 MeV from 2 keV. 
Since the required diffusion length logarithmically depends on the
energy gain $E/E_{\rm 0}$, we need the diffusion length of 270 km to
accelerate electrons to 10 MeV from 2 keV, 230 km to 50 MeV from 2 keV
for the same shock angle $\theta\sim 85$ degree. The acceleration to
ultra-high energy may be feasible, if we assume an energy-independent
diffusion length.

The entire acceleration time for $\theta\sim 80-85$ degree is
$L/2u\tan\theta \sim$ 0.6-0.3 s, which is consistent with the time
scale of the intensity fluctuation of the impulsive hard X-ray bursts,
satisfying the third requirement in section 2.1. The forth requirement
will be discussed in section 3.3.
  
\section{DISCUSSION AND SUMMARY}

Fermi acceleration has often been proposed to explain electron
acceleration in solar flares (eg. Bai 1983, Ellison \& Ramaty 1985,
Somov \& Kosugi 1997; first-order Fermi acceleration, Ramaty 1979,
LaRosa, Moore, \& Shore 1994; second-order Fermi acceleration). 
However, one of the principal problems associated with the first and
second-order Fermi acceleration is the high injection energy of about
20--100 keV (Bai 1983, Ramaty 1979). The oblique shock configuration
can drastically decrease the injection energy. Nevertheless, the
injection energy is still higher than the energy corresponding to the
coronal temperatures (1--3MK), and it may not be possible to
accelerate electrons from the coronal plasma, unless we assume a very
small diffusion length.  [There is no injection problem
with some versions of the stochastic acceleration (LaRosa,
Moore, \& Shore, 1994, and Miller, LaRosa, \& Moore, 1996).]

\subsection{New Observations from {\it\bf Yohkoh}}

Here, we point out that recent {\it Yohkoh} observations open up the
new possibility to accelerate electrons with first-order Fermi
acceleration: (1) Reconnection may produce a fast outflow with speed
exceeding the local sound speed. (The sound speed is much faster than
the Alfv\'en speed in the outflow.) Thus, it can produce a fast shock,
providing the site for first-order Fermi acceleration.  (2) Magnetic
field-line structure of the reconnection site naturally creates a
highly oblique shock. (The magnetic field lines are almost parallel to
the fast shock.)  (3) The outflow from the reconnection site is heated
upto 10--20MK by the slow shocks (Tsuneta 1996). Therefore, the plasma
temperature can reach the injection energy, and the bulk thermal
plasma can be accelerated.  (4) The acceleration site is bounded by
the two facing slow shocks, so that electrons that escape in the
upstream side during the course of acceleration most probably return
to the fast shock due to the magnetic mirror by the slow shocks,
subject to further acceleration. The electrons escaping in the
downstream side also returns the shock front, and continue to be
accelerated (Fig. 1).  Thus, the fast shock bounded by the two slow
shocks may provide an escape-free accelerator.

Indeed, we have shown that when the shock is oblique enough
($\theta\sim80-85$ degree), the electrons can be accelerated with
energy gain $E/E_{\rm 0} > 50$ in 0.3--0.6 s. The injection energy can
be low enough so that bulk acceleration from the thermal pool appears
to be feasible.  The diffusion length $l$ is safely as large as 600
km.  Although we have assumed an energy-independent diffusion, the
conclusion here is essentially the same as the energy-dependent case.

If the shock angle deviates to below about 80 degrees, then the
injection energy exceeds the thermal energy of the background plasma,
resulting in the significant reduction of the number of accelerated
electrons (unless we assume smaller diffusion length).  The shock
angle is not likely to be uniform due to non-uniform high-$\beta$
outflow in the upstream side of the fast shock (Tsuneta 1996). The
variety of flares from purely thermal flares to flares with highly
non-thermal nature may be related to this.

\subsection{Location of Hard X-ray Emitting Region}

Since there is no plasma density enhancement in the loop-top hard
X-ray source, the loop-top hard X-rays are essentially emitted from
the thick target process (Brown 1972). The column density $N$
[cm$^{-2}$] to stop electrons with energy $E$ [keV] is given by $ N =
8.3 \times 10^{17} E^2$. The column density required to stop $E$ = 50
keV electrons is $N \sim 2 \times 10^{21}$ cm$^{-2}$, and the
electrons have to bounce 100--200 times in the loop-top hard X-ray
source (Tsuneta et al. 1997, Wheatland \& Melrose 1995). It takes
about 5--10s to completely thermalize the accelerated electrons. This
dissipation time scale is much longer than the acceleration time scale
at the fast shock (0.3--0.6 s).  The loss cone angle for the slow
shocks is about 14 degrees for $B_{\rm u} = 52$G (field strength of
the upstream of the slow shock) and $B_{\rm d} = 3$ G (downstream of
the slow shock) (Tsuneta et al. 1997).  Thus, the slow shocks can
provide an efficient confinement mechanism.

These considerations indicate that the fast shock is located above
the loop-top hard X-ray source, and that the slow shock extends
downward below the fast shock to confine the non-thermal electrons.
The accelerated electrons are carried downward with slow outflow
downstream of the fast shock, and emit hard X-rays as long as they are
confined with the slow shock. They are no longer trapped outside the
slow shock region. The electrons are then immediately released toward
the footpoints, and may form the simultaneous double-source hard X-ray
structure at the footpoints of the reconnected field lines.

\subsection{Number of Accelerated Electrons}

Masuda (1994) reported that $2\times10^{35}$ electrons s$^{-1}$ are
accelerated during the peak phase of the 1992 January 13 flare (Masuda
1994). The reconnection outflow to the fast shock supplies thermal
electrons with rate $nuL^2 \sim 5\times10^{35}$ electrons s$^{-1}$, so
that the reconnection outflow can provide enough seed electrons to be
accelerated. Although acceleration efficiency is expected to be very
high as discussed above, we need more detailed modeling with Monte
Carlo simulation to obtain the spectra of the accelerated electrons,
and to answer whether the forth requirement described in section 2.1
is met.  This is left to our future work.

\acknowledgements
The authors thank H. Hudson for comments on the paper.  One of the
authors (TN) is indebted to the Research Fellowships of the Japan
Society for the Promotion of Science for Young Scientists.

\clearpage

\figcaption{
Magnetic field line configuration of the reconnection region. An
Alfv\'enic downward outflow is sandwiched by the two steady slow
shocks. A fast shock with length $L$ forms between the slow shocks. 
Magnetic disturbances both upstream and downstream of the fast shock
scatter the electrons being accelerated. The total length of the
diffusion region along the field lines is $l$.
\label{fig1}}

\figcaption{
The collisional energy loss rate and Fermi acceleration rate for 3
different shock angles. The net energy gain rate (thick lines) is the
energy gain rate (dotted and broken lines) minus the loss rate (thin
line).  
\label{fig2}}

\figcaption{
Eqs. (5) and (6) plotted on the plane of the diffusion length $l$ and
the shock angle $\theta$. The shock angle has to be $\sim$80--85
degree, and the diffusion length be $\sim$600 km for energy gain
$\sim$ 50 and $E_{\rm 0}=2$ keV.
\label{fig3}}

\clearpage

\plotone{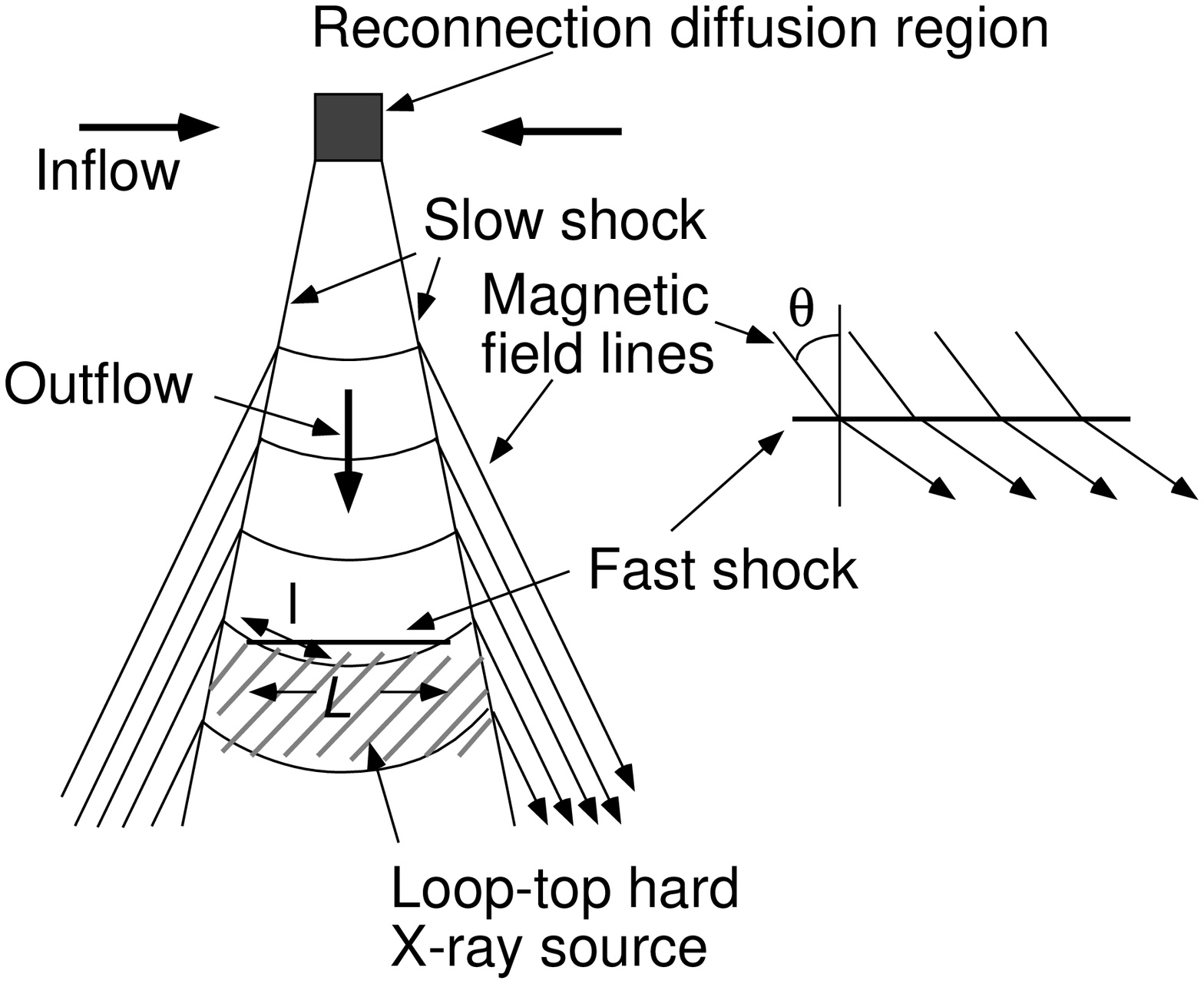}

\clearpage

\plotone{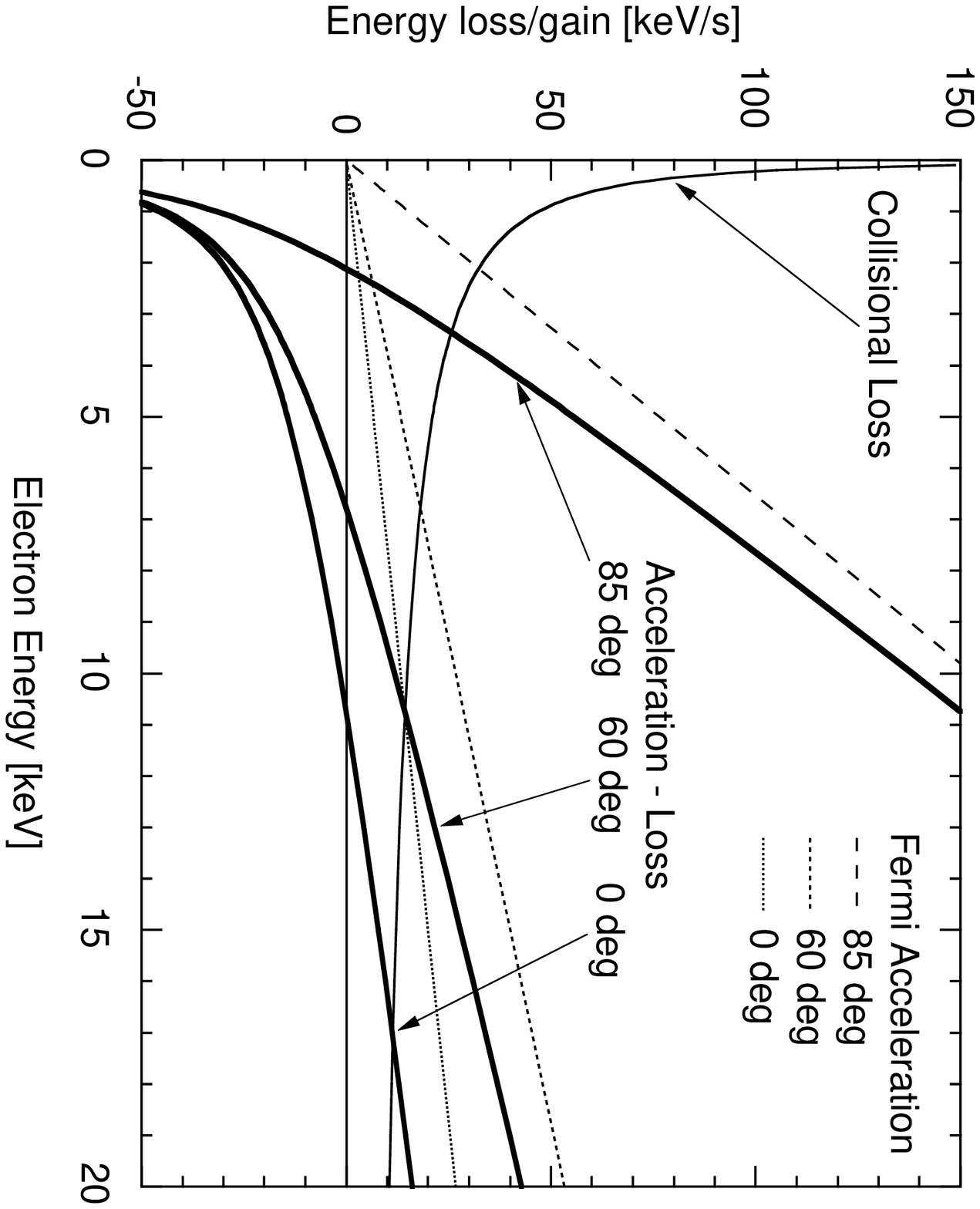}

\clearpage

\plotone{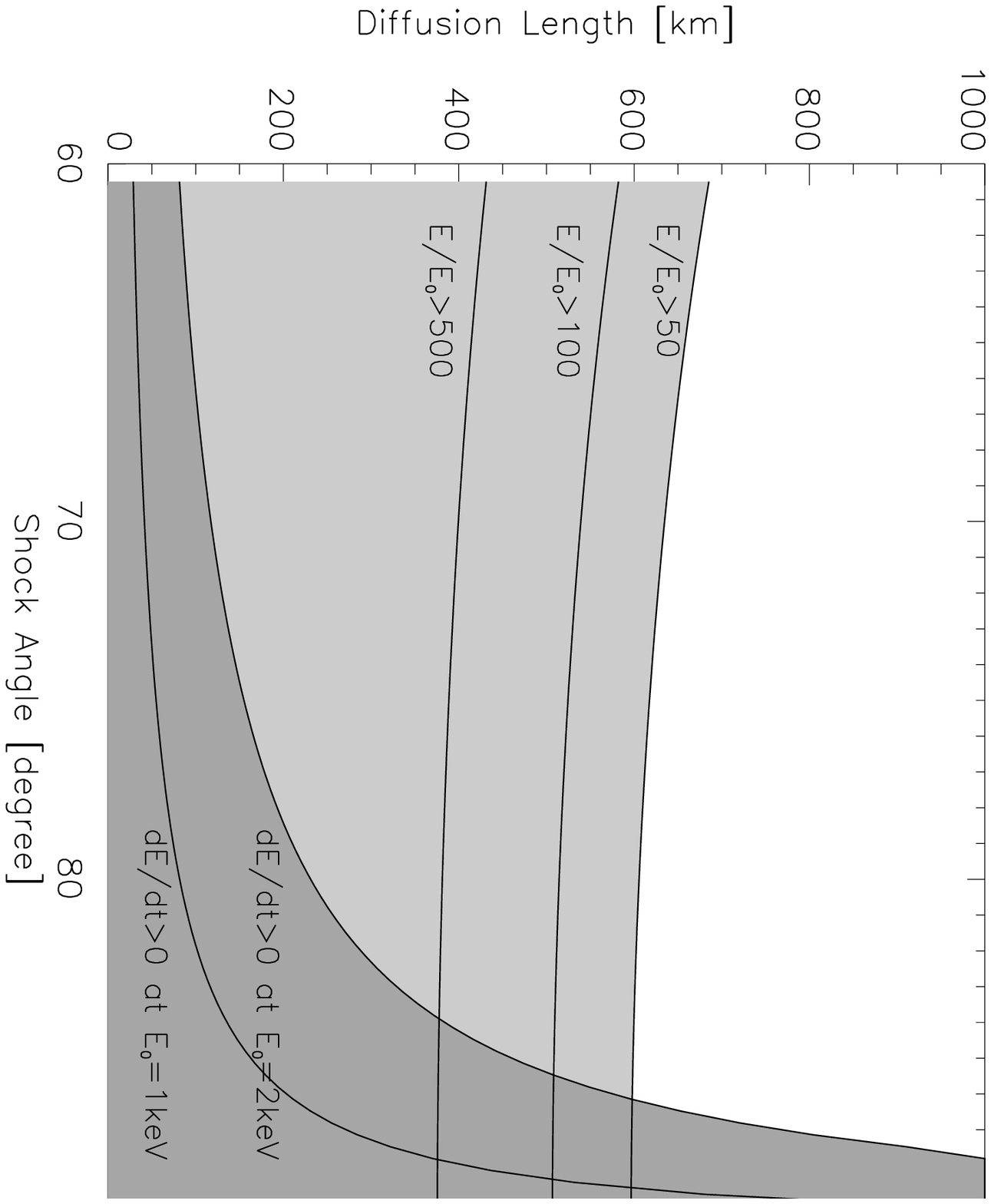}

\end{document}